\documentclass{mem}
\usepackage{natbib}\usepackage{txfonts}\usepackage{balance}
\usepackage{graphicx}
\usepackage[a4paper,breaklinks,dvipdfm]{hyperref}
\idline{84}{1}
\usepackage{txfonts} 

\def\cgs{erg~cm$^{-2}$~s$^{-1}$}
\def\cm{cm$^{-2}$}
\def\ergs{erg~s$^{-1}$}

\def\nh{{N$_{\rm H}$}}

\begin{document}
\def\teff{$T\rm_{eff }$}
\def\kms{$\mathrm {km s}^{-1}$}

\title{
The cosmic X-ray background: abundance and evolution of hidden black holes}

   \subtitle{}

\author{
R. \,Gilli 
          }

  \offprints{R.Gilli}
 
\institute{
INAF --
Osservatorio Astronomico di Bologna, Via Ranzani 1,
I-40127 Bologna, Italy\\
\email{roberto.gilli@oabo.inaf.it}
}

\authorrunning{Gilli}

\titlerunning{The XRB and hidden black holes}

\abstract{
The growth of supermassive black holes across cosmic time leaves a
radiative imprint recorded in the X-ray background (XRB). The
XRB spectral shape suggests that a large population of distant, hidden
nuclei must exist, which are now being revealed at higher and higher
redshifts by the deepest surveys performed by Chandra and XMM.

Our current understanding of the XRB emission in terms of AGN
population synthesis models is here reviewed, and the evolutionary
path of nuclear accretion and obscuration, as emerging from the major
X-ray surveys, is investigated. The role of galaxy merging versus secular
processes in triggering nuclear activity is also discussed in the
framework of recent galaxy/black hole co-evolutionary scenarios.
Finally, the limits of current instrumentation in the detection of the
most obscured and distant black holes are discussed and some possible
directions to overcome these limits are presented.
}
\maketitle{}

\begin{figure*}[t!]
\begin{center}
\resizebox{10cm}{!}{\includegraphics[clip=true,angle=270]{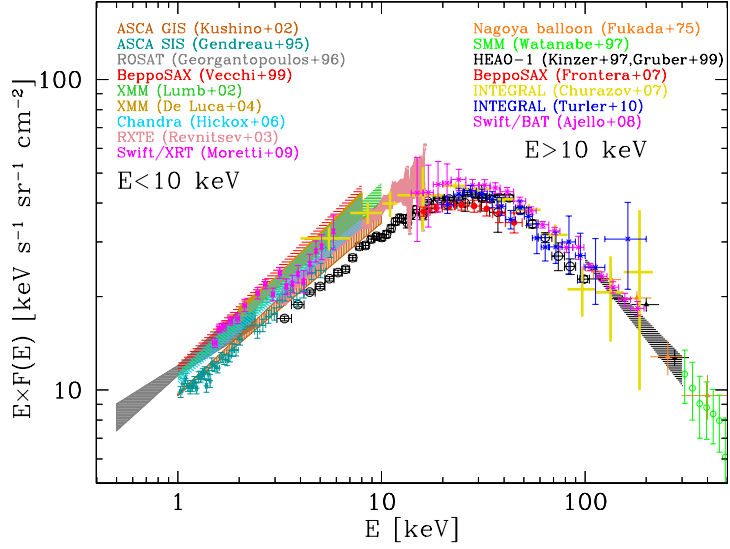}}
\caption{\footnotesize
Compilation of the measurements of the cosmic X-ray background spectrum in the 0.5 - 400 keV energy range.
Datapoints with different colors come from different combinations of missions and instruments as
labeled and referenced ({\it left labels}:  E$<$10 keV; {\it right labels}: E$>$10 keV).}
\label{xrb}
\end{center}
\end{figure*}

\section{Introduction}

Fifty years of technological progresses and scientific investigations
since its discovery, have revealed the main essence of the cosmic
X-ray background (XRB), the diffuse X-ray glow pervading the sky
discovered by Giacconi et al. in 1962 in the first X-ray astronomy
experiment.  From that first rocket flight to the deepest
observations now performed with Chandra and XMM, X-ray surveys gained
ten orders of magnitude in sensitivity. This progress showed that, at
energies below 10 keV, the XRB glow is resolved into hundreds of
millions of individual X-ray sources distributed across the whole
Universe: most of them are accreting supermassive black holes shining
as Active Galactic Nuclei (AGN) at different epochs. In the center of
the deepest X-ray field to date, the 4Ms Chandra Deep Field South (4Ms CDFS; Xue
et al. 2011), an AGN surface density of $\sim$$10^4$deg$^{-2}$ is
observed down to $f_{2-8keV}$$\sim5\times10^{-17}$erg cm$^{-2}$ s$^{-1}$,
whose integrated emission can virtually account for most of, if not
all, the XRB surface brigthness below 10 keV
\citep{xue11,moretti12}. At higher energies, however, in the 20-40 keV
energy range, where the XRB spectrum peaks, only a tiny fraction of
its emission has been resolved into individual sources, mainly because
of the limitations of high-energy instrumentation. A large portion of
the XRB emission at 30 keV is expected to be produced by the same
sources observed in the deep X-ray surveys like the 4Ms CDFS, just by
simply extrapolating their spectra to higher energies.  However, the
very shape of the XRB spectrum calls for an additional large
population of heavily obscured, Compton-thick AGN
(hereafter CT AGN, \nh$\gtrsim10^{24}$\cm) poorly sampled by most surveys at
E$<$10 keV. Despite several efforts, the cosmological evolution and
luminosity function of CT objects remain essentially
unknown and have to be postulated by AGN population synthesis models
trying to explain the XRB broad band spectrum (see next
Section). Heavily hidden black holes are then likely the key and
still missing ingredient to get a complete understanding of the cosmic
XRB, keeping it one of the most fascinating and
challenging topics in high-energy astrophysics.

\section{The synthesis of the X-ray background: Compton-thick AGN and
  the ``missing'' fraction}

The broad band, 0.5-400 keV XRB spectrum has a characteristic shape
peaking at E$\sim$30 keV. Below 10 keV it can be approximated by a
power law with $\Gamma$$\sim$1.4, i.e. it is harder than the average
spectrum of bright, unobscured QSOs. A compilation of XRB
measurements including the most recent results by Chandra and XMM at
E$<$10 keV, and BeppoSAX, Swift and INTEGRAL at E$>$10, keV is shown in
Fig.\ref{xrb}. The various measurements generally agree on the
spectral shape, but possess significant scatter, at the 20-30\%
level, in their absolute normalization (in general, most of the recent
XRB measures were found to be higher than the classic value measured
by HEAO-1 in the 80s). The origin of this scatter is still
debated. Part of it, but limited to the measures performed in pencil
beam surveys, could arise from cosmic variance. Another source of
uncertainty could be the stray-light which is entering an X-ray
telescope field of view if not properly modeled
\citep{moretti12}. Finally, $>$10\% calibration uncertainties have
been observed among different instruments on board different, or even
the same, missions \citep{tsu11}, which could explain another part of
the scatter. It is then fair to conclude that the XRB absolute flux is
known with a $\sim$20\% systematic uncertainty.
\begin{figure}[t!]
\resizebox{5.9cm}{!}{
\includegraphics[clip=true]{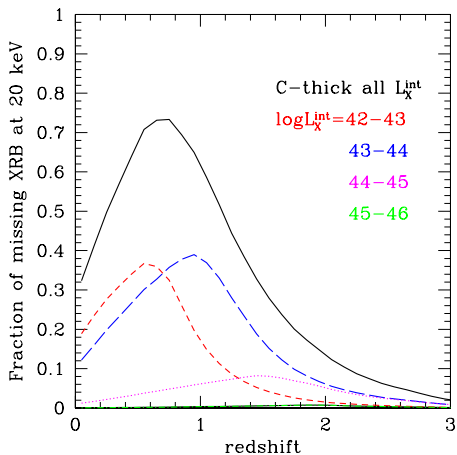}}
\caption{\footnotesize
Fraction of the XRB emission at 20 keV produced by CT AGN at different redshifts
in the \citet{gch07} model. The emission from the whole CT population (black line) is
splitted into four bins of intrinsic 2-10 keV luminosity
$L_x^{int}$ (color curves): most of the ``missing'' XRB is expected to be produced by objects with $L_x^{int}<10^{44}$ erg s$^{-1}$ and $z<1$.}
\label{zdist}
\end{figure}

In 1989 Setti \& Woltjer first proposed that the hard XRB spectrum
could be explained by the superposition of AGN spectra with different
absorption degrees. This prompted the flourishing of a series of AGN
population synthesis models with an increasing degree of complexity,
including: detailed spectra and absorption distributions
\citep{madau94,comastri95}; a careful treatment of Compton scattering
in heavily obscured sources \citep{wilman99,pompilio00}; the inclusion
of iron features \citep{gilli99}; evolution of the obscured AGN
fraction with luminosity \citep{ueda03} and redshift
\citep{lafranca05,balla06,tuv09}; dispersion in the AGN primary
spectral slopes \citep{gch07}; exploration of the full
parameter space (e.g. reflection fraction vs CT abundance;
\citealt{tuv09,akylas12}).  These works suggested that 10-30\% of the
XRB emission at 30 keV, depending on the model assumptions and on the
adopted XRB normalization, is produced by CT AGN. \citet{gch07}
splitted the population of CT AGN into a pair of evenly populated
categories of transmission-dominated (\nh$\sim$$10^{24}$\cm) and
reflection dominated (\nh$>$$10^{25}$\cm) objects, and built a
self-consistent synthesis model in which CT AGN are as abundant as less absorbed
objects. However, this result is far from being conclusive: as
\citet{comastri08} showed, varying the number of reflection-dominated
objects by a factor of four was still consistent with the XRB
spectrum. This suggests that the XRB spectrum has limited constraining
power and that an accurate estimate of the abundance of CT AGN can
only be determined by $direct$ object selection.

\begin{figure}[t!]
\resizebox{5.9cm}{!}{
\includegraphics[clip=true]{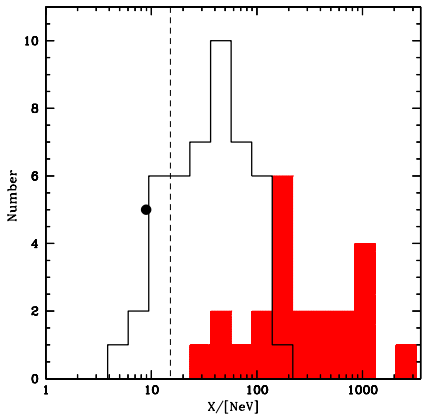}}
\caption{\footnotesize
The distribution of the observed 2-10 keV to [Ne~V] luminosity ratio (X/NeV) for the [Ne~V]-selected Type-2 AGN in COSMOS.
Filled and empty histograms refer to  X-ray detections and upper limits, respectively. The filled dot (X/NeV$\sim$9, arbitrary position on the $y$-axis) 
is derived by stacking X-ray undetected sources. The vertical dashed line marks the threshold to select candidate
CT AGN (X/NeV$<$15). From \citet{mignoli13}.}
\label{neon}
\end{figure}
In Fig.\ref{zdist} it is shown the fraction of the XRB emission at 20
keV produced by CT AGN (the so-called``missing'' XRB) as a function of
their redshift in the \citet{gch07} model. Most of the ``missing''
XRB is expected to be produced by Seyfert-like CT objects at
$z$$\sim$0.7 (i.e. with intrinsic 2-10 keV luminosity
$L_x^{int}$$<$$10^{44}$ erg s$^{-1}$). A direct sampling of this very
population is therefore highly desirable, and it has been recently
attempted by \citet{mignoli13} who used the [Ne~V]3426\AA\ emission
line to pinpoint narrow line (type-2) AGN among the $\sim$7400
galaxies at 0.65$<$z$<$1.2 in the zCOSMOS spectroscopic survey
\citep{lilly07}. Seventytwo type-2 AGN with Seyfert-like luminosities
and average redshift 0.9 were found in the survey area covered by
Chandra ($\sim$0.9deg$^2$, the C-COSMOS survey, \citealt{elvis09}):
among them, 46 objects with X-ray to NeV ratio lower than 100 were
accounted as heavily obscured AGN, including 9 CT candidates with
X/NeV$<15$. The distribution of the X/NeV ratios for this sample is shown in Fig.\ref{neon}: 
by stacking the X-ray undetected sources we derived an average X/NeV ratio of 9, which points
towards a CT fraction in the whole type-2 sample of $\approx41\%$ (see the 
detailed simulations in \citealt{mignoli13}), in line with the 50\% fraction assumed in
the \cite{gch07} model. This measurement will be further refined by the forthcoming 
2.8Ms Chandra COSMOS-Legacy data (PI F. Civano).
 
\section{Evolution of nuclear obscuration and triggering mechanisms}

While an overall self-consistent picture is emerging on the CT AGN
population from the previous estimates, it is recalled that any
estimate of nuclear absorption based on the comparison between some
proxy of the intrinsic power (e.g.  high-ionization lines such as 
[O~III] or [Ne~V], or mid-infrared luminosity) and the observed X-ray
emission is highly uncertain, and in turn affects the estimate of the
space density of CT AGN (see
e.g. \citealt{fiore09,v10,geo11,alex11}). The only unambiguous way to
select bona-fide CT AGN is through X-ray spectroscopy, and deep X-ray
surveys are key to this. As an example, several CT AGN, up to
$z$$\sim$5, have been reported in the CDFS based either on Chandra or
XMM data \citep{norman02,tozzi06,geo07,comastri11,feruglio11,gilli11}.
These numbers are too small to compute their space density and
evolution, and this situation is unlike to change unless new X-ray
facilities become available (see next Section).  Nonetheless, X-ray
spectroscopy in the deep Chandra and XMM surveys is shedding light on
another very controversial issue, i.e. the evolution history of
Compton-thin obscuration. Several works in the past years claimed that
the fraction of obscured nuclei increases with redshift
\citep{lafranca05,h08,tuv09}. This relation was not found by others
\citep{ueda03,gch07}, leading \citet{gilli10} to propose that strong
selection effects were producing most of it. X-ray spectroscopy in the
Chandra/XMM deep fields is shedding new light on this
issue. \citet{iwasawa12} analyzed the XMM spectra of 46 AGN at $z>1.7$
in the XMM-CDFS and found that the fraction of luminous
($L_{2-10}^{rest}\gtrsim10^{44}$\ergs) obscured QSOs was significantly
higher than that observed in the local Universe by \citet{burlon11},
and that selection biases could not explain this difference. Based on
the Chandra spectra of a sample of 34 AGN at $z>3$ in the 4Ms CDFS,
\citet{vito13} found that the increase of the obscured AGN fraction
with redshift was less significant or even absent at lower
luminosities. This behaviour has an interesting qualitative
interpretation in a scenario involving two modes of
accretion/obscuration, which could mirror the two modes of star
formation (bursting vs secular/main sequence) observed in the galaxy
population \citep{elbaz11,rodighiero11}. Indeed, bright QSOs are
preferentially found in starburst systems, while lower-luminosity
objects are preferentially seen in main sequence galaxies
\citep{rovilos12}. Furthermore, the fraction of AGN and star forming
systems in mergers increases with their bolometric luminosity and
``starburstiness'', respectively
\citep{treister12,kartaltepe12}. Based on all this evidence,
\citet{iwasawa12} suggested that the luminosity-dependent increase of
the obscured AGN fraction with redshift could be related to the higher
merger rate $and$ gas fraction $f_{gas}$ in galaxies at high redshift. As
often postulated (e.g. \citealt{hop08}), mergers/interactions could
trigger both powerful nuclear accretion and a starburst with compact
and chaotic geometry, installing an evolutionary sequence in which,
after an initial obscured phase, the black holes quickly (0.01-0.1
Gyr) clears its close environment and shines as an unobscured QSO. The
higher merger rate alone, however, would simply increase the abundance
of the $total$ QSO population at high-z. It is the higher gas fraction
that would produce a longer obscured phase at high-z and hence a
higher fraction of obscured QSOs. If lower-luminosity objects are
instead preferentially hosted by main sequence, non-interacting
galaxies, the overall system geometry could be regular. Hence, a
static geometry in the nuclear regions, as postulated in the classic Unified
Models, could be valid and the observed obscuration would only depend
on the line of sight. In this case, although a larger gas reservoir
(higher $f_{gas}$) is available at high-z, the lack of interactions
would keep it away from the BH: the available gas would then just be
consumed to form stars. The system geometry will then be preserved in
time and, in turn, the observed fraction of obscured AGN. This
appealing scenario admittedly relies on a qualitative interpretation:
further observations and detailed modeling are needed to
test it in the next few years.
\begin{figure}[t!]
\resizebox{6.1cm}{!}{\includegraphics[clip=true,angle=270]{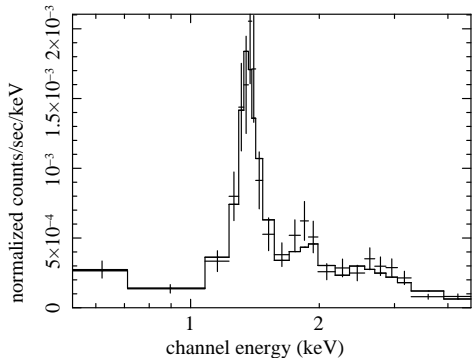}}
\caption{\footnotesize
Simulated WFXT spectrum of CDFS-202, a CT AGN at z=3.7 \citep{norman02,comastri11}. About 500
net counts are collected in 400 ksec (the exposure of the WFXT deep survey).  The iron line 
is clearly detected and allows an accurate redshift determination.}
\label{202}
\end{figure}

\section{Prospects for the very near and mid-term future}

Besides further analysis of Chandra and XMM data from completed (CDFS,
COSMOS, AEGIS, XBootes) and on-going Ms surveys (COSMOS-Legacy, XXL),
a fundamental contribution to our understanding of hidden black holes
is shortly expected by NuSTAR \citep{harrison10}, on orbit since June
2012.  NuSTAR is carrying the first focusing optics at E$>$10 keV and
will overtake the sensitivity of current high-energy instruments by
more than two orders of magnitude. The NuSTAR surveys in the CDFS and
COSMOS fields will likely reach sensitivities of a few
$\times10^{-14}$\cgs in the 10-40 keV range.  These appear obviously
very shallow if compared with the Chandra or XMM data in the same
fields. It is then unlikely that NuSTAR uncovers new X-ray sources in
the CDFS and COSMOS, but high-energy data will significantly improve
the $N_H$ estimate for those already known. As far as the demography
of CT AGN is concerned, the whole NuSTAR program is expected
to detect Seyfert-like objects ($L_x$$<$$10^{44}$erg~s$^{-1}$) up to
z$<$0.5 (in small numbers though), opening a new territory to test the
assumptions of XRB synthesis models.

\begin{figure}[t!]
\resizebox{6.3cm}{!}{\includegraphics[clip=true]{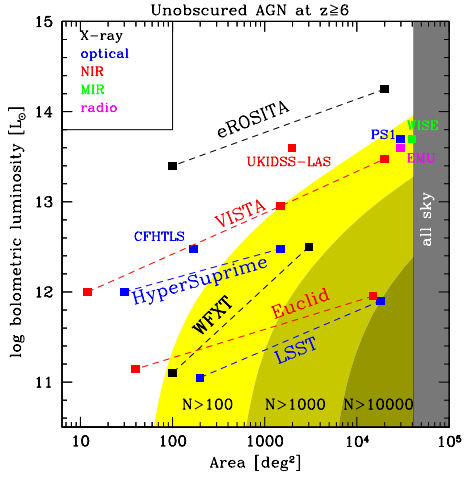}}
\caption{\footnotesize
Minimum bolometric luminosity of unobscured AGN at $z\geq6$ detectable over a given survey area. 
Different colors refer to surveys in different wavebands  (see upper-left corner).
Shaded areas with progressively darker colors show the regions 
where a given survey should detect more than 100, 1000, and 10000 objects (assuming an exponential decline
of the space density of AGN at $z>$3; \citealt{brusa09,civano11}. WFXT matches in area and sensitivity future optical (HyperSuprime and LSST) and near-IR (Euclid) surveys.}
\label{z6}
\end{figure}

A real breakthrough in the study of the demographics of CT AGN would
be possible in the next years if concept-missions like the Wide Field
X-ray Telescope (WFXT; see Murray et al. this volume) will come into
reality. WFXT would produce wide-and-deep surveys in the 0.5-7 keV
band with an unprecedented grasp and $average$ resolution across the
field of view comparable to that of Chandra.  Unlike standard
Wolter-type mirrors, the polynomial mirrors desing of WFXT is
optimized to provide a constant PSF with off-axis angle, making this
instrument a perfect survey machine. To give a sense of its
capabilities, with its current survey strategy WFXT is expected to
produce the equivalent of $\sim$1000 CDFS surveys (at the 1Ms CDFS
depth) $plus$ 3000 C-COSMOS surveys. Based on the predictions of
current XRB synthesis models, WFXT is expected to return $\sim$ten
millions of AGN up to the highest redshifts ($z>6$): about
$5\times10^5$ AGN will be detected with more than 400 net counts,
including $10^4$ objects with $N_H>10^{23}$\cm, for which accurate
X-ray spectra will be obtained. This spectral sample will allow an
unambiguous determination of how unobscured and moderately obscured
AGN evolve as a function of redshift, luminosity, column density and
environment, providing key information on BH/galaxy coevolution
theories. In this sample, about 300 objects are expected to be {\it
  bona-fide} CT AGN at $z>1$. As shown by the simulation in
Fig.\ref{202}, their WFXT spectra will reveal $directly$ their CT
nature, and measure their redshift through the detection of the iron
K$\alpha$ feature. The cosmological evolution of the elusive CT AGN
population will then be probed for the first time. Finally, WFXT is
expected to detect $>1000$ obscured and unobscured AGN at $z>6$
(current model predictions vary by orders of magnitude), hence
determining how the first black holes evolve: as shown in
Fig.\ref{z6}, these data will perfectly match the other wide-and-deep
multiwavelength survey data available in a 5-10 years
time frame.

\begin{acknowledgements}
 
I wish to thank the members of the CDFS, COSMOS and WFXT
collaborations for their help, and the organizers
of the conference ``X-ray astronomy: towards the next 50 years'' for
their kind invitation. Support from ASI-INAF grant I/009/10/0 and
PRIN-INAF 2011 is acknowledged.

\end{acknowledgements}

\begin{thebibliography}{45}
\expandafter\ifx\csname natexlab\endcsname\relax\def\natexlab#1{#1}\fi

\bibitem[{{Akylas} {et~al.}(2012)}]{akylas12}{Akylas}, A., {et~al.} 2012, \aap, 546, A98

\bibitem[{{Alexander} {et~al.}(2011)}]{alex11}
{Alexander}, D.~M., {et~al.} 2011, \apj, 738,  44

\bibitem[{{Ballantyne} {et~al.}(2006){Ballantyne}, {Everett}, \&
  {Murray}}]{balla06}
{Ballantyne}, D.~R.,  {et~al.} 2006, \apj, 639, 740

\bibitem[{{Brusa} {et~al.}(2009)}]{brusa09}
{Brusa}, M., {et~al.} 2009, \apj, 693, 8

\bibitem[{{Burlon} {et~al.}(2011){Burlon}, {Ajello}, {Greiner}, {Comastri},
  {Merloni}, \& {Gehrels}}]{burlon11}
{Burlon}, D., {et~al.} 2011, \apj, 728, 58

\bibitem[{{Civano} {et~al.}(2011){Civano}, {Brusa}, {Comastri}, {Elvis},
  {Salvato}, {Zamorani}, {Capak}, {Fiore}, {Gilli}, {Hao}, {Ikeda}, {Kakazu},
  {Kartaltepe}, {Masters}, {Miyaji}, {Mignoli}, {Puccetti}, {Shankar},
  {Silverman}, {Vignali}, {Zezas}, \& {Koekemoer}}]{civano11}
{Civano}, F.,  {et~al.} 2011, \apj, 741, 91

\bibitem[{{Comastri} {et~al.}(2008){Comastri}, {Gilli}, {Fiore}, {Vignali},
  {Della Ceca}, \& {Malaguti}}]{comastri08}
{Comastri}, A., {et~al.} 2008, MmSAI, 79, 59

\bibitem[{{Comastri} {et~al.}(2011){Comastri}, {Ranalli}, {Iwasawa}, {Vignali},
  {Gilli}, {Georgantopoulos}, {Barcons}, {Brandt}, {Brunner}, {Brusa},
  {Cappelluti}, {Carrera}, {Civano}, {Fiore}, {Hasinger}, {Mainieri},
  {Merloni}, {Nicastro}, {Paolillo}, {Puccetti}, {Rosati}, {Silverman},
  {Tozzi}, {Zamorani}, {Balestra}, {Bauer}, {Luo}, \& {Xue}}]{comastri11}
{Comastri}, A., {et~al.} 2011, \aap, 526, L9+

\bibitem[{{Comastri} {et~al.}(1995){Comastri}, {Setti}, {Zamorani}, \&
  {Hasinger}}]{comastri95}
{Comastri}, A., {et~al.}  1995, \aap, 296,
  1

\bibitem[{{Elbaz} {et~al.}(2011){Elbaz}, {Dickinson}, {Hwang},
  {D{\'{\i}}az-Santos}, {Magdis}, {Magnelli}, {Le Borgne}, {Galliano},
  {Pannella}, {Chanial}, {Armus}, {Charmandaris}, {Daddi}, {Aussel}, {Popesso},
  {Kartaltepe}, {Altieri}, {Valtchanov}, {Coia}, {Dannerbauer}, {Dasyra},
  {Leiton}, {Mazzarella}, {Alexander}, {Buat}, {Burgarella}, {Chary}, {Gilli},
  {Ivison}, {Juneau}, {Le Floc'h}, {Lutz}, {Morrison}, {Mullaney}, {Murphy},
  {Pope}, {Scott}, {Brodwin}, {Calzetti}, {Cesarsky}, {Charlot}, {Dole},
  {Eisenhardt}, {Ferguson}, {F{\"o}rster Schreiber}, {Frayer}, {Giavalisco},
  {Huynh}, {Koekemoer}, {Papovich}, {Reddy}, {Surace}, {Teplitz}, {Yun}, \&
  {Wilson}}]{elbaz11}
{Elbaz}, D., {et~al.} 2011, \aap, 533, A119

\bibitem[{{Elvis} {et~al.}(2009){Elvis}, {Civano}, {Vignali}, {Puccetti},
  {Fiore}, {Cappelluti}, {Aldcroft}, {Fruscione}, {Zamorani}, {Comastri},
  {Brusa}, {Gilli}, {Miyaji}, {Damiani}, {Koekemoer}, {Finoguenov}, {Brunner},
  {Urry}, {Silverman}, {Mainieri}, {Hasinger}, {Griffiths}, {Carollo}, {Hao},
  {Guzzo}, {Blain}, {Calzetti}, {Carilli}, {Capak}, {Ettori}, {Fabbiano},
  {Impey}, {Lilly}, {Mobasher}, {Rich}, {Salvato}, {Sanders}, {Schinnerer},
  {Scoville}, {Shopbell}, {Taylor}, {Taniguchi}, \& {Volonteri}}]{elvis09}
{Elvis}, M., {et~al.} 2009, \apjs, 184, 158

\bibitem[{{Feruglio} {et~al.}(2011){Feruglio}, {Daddi}, {Fiore}, {Alexander},
  {Piconcelli}, \& {Malacaria}}]{feruglio11}
{Feruglio}, C., {et~al.} 2011, \apjl, 729, L4+

\bibitem[{{Fiore} {et~al.}(2009){Fiore}, {Puccetti}, {Brusa}, {Salvato},
  {Zamorani}, {Aldcroft}, {Aussel}, {Brunner}, {Capak}, {Cappelluti}, {Civano},
  {Comastri}, {Elvis}, {Feruglio}, {Finoguenov}, {Fruscione}, {Gilli},
  {Hasinger}, {Koekemoer}, {Kartaltepe}, {Ilbert}, {Impey}, {LeFloc'h},
  {Lilly}, {Mainieri}, {Martinez-Sansigre}, {McCracken}, {Menci}, {Merloni},
  {Miyaji}, {Sanders}, {Sargent}, {Schinnerer}, {Scoville}, {Silverman},
  {Smolcic}, {Steffen}, {Santini}, {Taniguchi}, {Thompson}, {Trump}, {Vignali},
  {Urry}, \& {Yan}}]{fiore09}
{Fiore}, F., {et~al.} 2009, \apj, 693, 447

\bibitem[{{Georgantopoulos} {et~al.}(2007){Georgantopoulos}, {Georgakakis}, \&
  {Akylas}}]{geo07}
{Georgantopoulos},I.,{et al.} 2007,\aap,466,823

\bibitem[{{Georgantopoulos} {et~al.}(2011){Georgantopoulos}, {Rovilos},
  {Akylas}, {Comastri}, {Ranalli}, {Vignali}, {Balestra}, {Gilli}, \&
  {Cappelluti}}]{geo11}
{Georgantopoulos}, I.,{et~al.} 2011,\aap,534,A23

\bibitem[{{Giacconi} {et~al.}(1962){Giacconi}, {Gurski}, {Paolini}, \&
  {Rossi}}]{giacconi62}
{Giacconi}, R.,{et~al.} 1962, Phys. Rev. Lett. 9, 439


\bibitem[{{Gilli} {et~al.}(1999){Gilli}, {Comastri}, {Brunetti}, \&
  {Setti}}]{gilli99}
{Gilli}, R., {et~al.} 1999, New  Astronomy, 4, 45

\bibitem[{{Gilli} {et~al.}(2007){Gilli}, {Comastri}, \& {Hasinger}}]{gch07}
{Gilli}, R., {et~al.} 2007, \aap, 463, 79

\bibitem[{{Gilli} {et~al.}(2010){Gilli}, {Comastri}, {Vignali},
  {Ranalli}, \& {Iwasawa}}]{gilli10}
{Gilli}, R., {et~al.}
  2010{\natexlab{a}}, AIPC, 1248, 359

\bibitem[{{Gilli} {et~al.}(2011){Gilli}, {Su}, {Norman}, {Vignali}, {Comastri},
  {Tozzi}, {Rosati}, {Stiavelli}, {Brandt}, {Xue}, {Luo}, {Castellano},
  {Fontana}, {Fiore}, {Mainieri}, \& {Ptak}}]{gilli11}
{Gilli}, R., {et~al.} 2011, \apjl, 730, L28

\bibitem[{{Harrison} {et~al.}(2010){Harrison}, {Boggs}, {Christensen}, {Craig},
  {Hailey}, {Stern}, {Zhang}, {Angelini}, {An}, {Bhalereo}, {Brejnholt},
  {Cominsky}, {Cook}, {Doll}, {Giommi}, {Grefenstette}, {Hornstrup}, {Kaspi},
  {Kim}, {Kitaguchi}, {Koglin}, {Liebe}, {Madejski}, {Kruse Madsen}, {Mao},
  {Meier}, {Miyasaka}, {Mori}, {Perri}, {Pivovaroff}, {Puccetti}, {Rana}, \&
  {Zoglauer}}]{harrison10}
{Harrison}, F.~A., {et~al.} 2010, SPIE, 7732

\bibitem[{{Hasinger}(2008)}]{h08}
{Hasinger}, G. 2008, \aap, 490, 905


\bibitem[{{Hopkins} {et~al.}(2008){Hopkins}, {Hernquist}, {Cox}, \& {Kere{\v
  s}}}]{hop08}
{Hopkins}, P.~F., {et~al.} 2008,
  \apjs, 175, 356

\bibitem[{{Iwasawa} {et~al.}(2012){Iwasawa}, {Gilli}, {Vignali}, {Comastri},
  {Brandt}, {Ranalli}, {Vito}, {Cappelluti}, {Carrera}, {Falocco},
  {Georgantopoulos}, {Mainieri}, \& {Paolillo}}]{iwasawa12}
{Iwasawa}, K., {et~al.} 2012, \aap, 546, A84

\bibitem[{{Kartaltepe} {et~al.}(2012){Kartaltepe}, {Dickinson}, {Alexander},
  {Bell}, {Dahlen}, {Elbaz}, {Faber}, {Lotz}, {McIntosh}, {Wiklind}, {Altieri},
  {Aussel}, {Bethermin}, {Bournaud}, {Charmandaris}, {Conselice}, {Cooray},
  {Dannerbauer}, {Dav{\'e}}, {Dunlop}, {Dekel}, {Ferguson}, {Grogin}, {Hwang},
  {Ivison}, {Kocevski}, {Koekemoer}, {Koo}, {Lai}, {Leiton}, {Lucas}, {Lutz},
  {Magdis}, {Magnelli}, {Morrison}, {Mozena}, {Mullaney}, {Newman}, {Pope},
  {Popesso}, {van der Wel}, {Weiner}, \& {Wuyts}}]{kartaltepe12}
{Kartaltepe}, J.~S., {et~al.} 2012, \apj,  757, 23

\bibitem[{{La Franca} {et~al.}(2005){La Franca}, {Fiore}, {Comastri}, {Perola},
  {Sacchi}, {Brusa}, {Cocchia}, {Feruglio}, {Matt}, {Vignali}, {Carangelo},
  {Ciliegi}, {Lamastra}, {Maiolino}, {Mignoli}, {Molendi}, \&
  {Puccetti}}]{lafranca05}
{La Franca}, F., {et~al.} 2005, \apj, 635, 864

\bibitem[{{Lilly} {et~al.}(2007){Lilly}, {Le F{\`e}vre}, {Renzini}, {Zamorani},
  {Scodeggio}, {Contini}, {Carollo}, {Hasinger}, {Kneib}, {Iovino}, {Le Brun},
  {Maier}, {Mainieri}, {Mignoli}, {Silverman}, {Tasca}, {Bolzonella},
  {Bongiorno}, {Bottini}, {Capak}, {Caputi}, {Cimatti}, {Cucciati}, {Daddi},
  {Feldmann}, {Franzetti}, {Garilli}, {Guzzo}, {Ilbert}, {Kampczyk}, {Kovac},
  {Lamareille}, {Leauthaud}, {Borgne}, {McCracken}, {Marinoni}, {Pello},
  {Ricciardelli}, {Scarlata}, {Vergani}, {Sanders}, {Schinnerer}, {Scoville},
  {Taniguchi}, {Arnouts}, {Aussel}, {Bardelli}, {Brusa}, {Cappi}, {Ciliegi},
  {Finoguenov}, {Foucaud}, {Franceschini}, {Halliday}, {Impey}, {Knobel},
  {Koekemoer}, {Kurk}, {Maccagni}, {Maddox}, {Marano}, {Marconi}, {Meneux},
  {Mobasher}, {Moreau}, {Peacock}, {Porciani}, {Pozzetti}, {Scaramella},
  {Schiminovich}, {Shopbell}, {Smail}, {Thompson}, {Tresse}, {Vettolani},
  {Zanichelli}, \& {Zucca}}]{lilly07}
{Lilly}, S.~J.,  {et~al.} 2007, \apjs, 172,  70

\bibitem[{{Madau} {et~al.}(1994){Madau}, {Ghisellini}, \& {Fabian}}]{madau94}
{Madau}, P., {et~al.} 1994, \mnras, 270, L17

\bibitem[{{Mignoli et al.}(2013)}]{mignoli13}
{Mignoli et al.}, M. 2013, \aap,~submitted

\bibitem[{{Moretti} {et~al.}(2012){Moretti}, {Vattakunnel}, {Tozzi},
  {Salvaterra}, {Severgnini}, {Fugazza}, {Haardt}, \& {Gilli}}]{moretti12}
{Moretti}, A., {et~al.} 2012, \aap, 548, A87

\bibitem[{{Norman} {et~al.}(2002){Norman}, {Hasinger}, {Giacconi}, {Gilli},
  {Kewley}, {Nonino}, {Rosati}, {Szokoly}, {Tozzi}, {Wang}, {Zheng}, {Zirm},
  {Bergeron}, {Gilmozzi}, {Grogin}, {Koekemoer}, \& {Schreier}}]{norman02}
{Norman}, C., {et~al.} 2002, \apj, 571, 218

\bibitem[{{Pompilio} {et~al.}(2000){Pompilio}, {La Franca}, \&
  {Matt}}]{pompilio00}
{Pompilio}, F., {et~al.} 2000, \aap, 353, 440

\bibitem[{{Rodighiero} {et~al.}(2011){Rodighiero}, {Daddi}, {Baronchelli},
  {Cimatti}, {Renzini}, {Aussel}, {Popesso}, {Lutz}, {Andreani}, {Berta},
  {Cava}, {Elbaz}, {Feltre}, {Fontana}, {F{\"o}rster Schreiber},
  {Franceschini}, {Genzel}, {Grazian}, {Gruppioni}, {Ilbert}, {Le Floch},
  {Magdis}, {Magliocchetti}, {Magnelli}, {Maiolino}, {McCracken}, {Nordon},
  {Poglitsch}, {Santini}, {Pozzi}, {Riguccini}, {Tacconi}, {Wuyts}, \&
  {Zamorani}}]{rodighiero11}
{Rodighiero}, G., {et~al.} 2011, \apjl, 739,  L40


\bibitem[{{Rovilos} {et~al.}(2012){Rovilos}, {Comastri}, {Gilli},
  {Georgantopoulos}, {Ranalli}, {Vignali}, {Lusso}, {Cappelluti}, {Zamorani},
  {Elbaz}, {Dickinson}, {Hwang}, {Charmandaris}, {Ivison}, {Merloni}, {Daddi},
  {Carrera}, {Brandt}, {Mullaney}, {Scott}, {Alexander}, {Del Moro},
  {Morrison}, {Murphy}, {Altieri}, {Aussel}, {Dannerbauer}, {Kartaltepe},
  {Leiton}, {Magdis}, {Magnelli}, {Popesso}, \& {Valtchanov}}]{rovilos12}
{Rovilos}, E., {et~al.} 2012, \aap, 546, A58

\bibitem[{{Setti} \& {Woltjer}(2006)}]{sw89}
{Setti}, G. \& {Woltjer}, L. 1989, \aap, 224, L21

\bibitem[{{Tozzi} {et~al.}(2006){Tozzi}, {Gilli}, {Mainieri}, {Norman},
  {Risaliti}, {Rosati}, {Bergeron}, {Borgani}, {Giacconi}, {Hasinger},
  {Nonino}, {Streblyanska}, {Szokoly}, {Wang}, \& {Zheng}}]{tozzi06}
{Tozzi}, P., {et~al.} 2006, \aap, 451, 457

\bibitem[{{Treister} {et~al.}(2012){Treister}, {Schawinski}, {Urry}, \&
  {Simmons}}]{treister12}
{Treister}, E.,  {et~al.} 2012,
  \apjl, 758, L39

\bibitem[{{Treister} {et~al.}(2009){Treister}, {Urry}, \& {Virani}}]{tuv09}
{Treister}, E.,  {et~al.} 2009, \apj, 696, 110

\bibitem[{{Tsujimoto} {et~al.}(2011){Tsujimoto}}]{tsu11}
{Tsujimoto}, M.,  {et~al.} 2011, \aap, 525, A25

\bibitem[{{Ueda} {et~al.}(2003){Ueda}, {Akiyama}, {Ohta}, \& {Miyaji}}]{ueda03}
{Ueda}, Y.,  {et~al.} 2003, \apj, 598, 886

\bibitem[{{Vignali} {et~al.}(2010){Vignali}, {Alexander}, {Gilli}, \&
  {Pozzi}}]{v10}
{Vignali}, C.,  {et~al.} 2010, \mnras,
  404, 48

\bibitem[{{Vito} {et~al.}(2013){Vito}, {Vignali}, {Gilli}, {Comastri},
  {Iwasawa}, {Brandt}, {Alexander}, {Brusa}, {Lehmer}, {Bauer}, {Schneider},
  {Xue}, \& {Luo}}]{vito13}
{Vito}, F., {et~al.} 2013, \mnras, 428, 354

\bibitem[{{Wilman} \& {Fabian}(1999)}]{wilman99}
{Wilman}, R.~J. {et~al.}  1999, \mnras, 309, 862

\bibitem[{{Xue} {et~al.}(2011){Xue}, {Luo}, {Brandt}, {Bauer}, {Lehmer},
  {Broos}, {Schneider}, {Alexander}, {Brusa}, {Comastri}, {Fabian}, {Gilli},
  {Hasinger}, {Hornschemeier}, {Koekemoer}, {Liu}, {Mainieri}, {Paolillo},
  {Rafferty}, {Rosati}, {Shemmer}, {Silverman}, {Smail}, {Tozzi}, \&
  {Vignali}}]{xue11}
{Xue}, Y.~Q.,  {et~al.} 2011, \apjs, 195, 10

\end{thebibliography}

\end{document}